\documentstyle[preprint,tighten,aps]{revtex}

\def\Sppzero{S}

\def\lapprox{\mathrel{\mathop
  {\hbox{\lower0.5ex\hbox{$\sim$}\kern-0.8em\lower-0.7ex\hbox{$<$}}}}}
\def\gapprox{\mathrel{\mathop
  {\hbox{\lower0.5ex\hbox{$\sim$}\kern-0.8em\lower-0.7ex\hbox{$>$}}}}}

\begin{document}

\preprint{\vbox{\noindent
          \null\hfill  INFNFE-06-97}}
\title{Helioseismology and  p+p $\rightarrow$ d + e$^+$ + $\nu _e$ 
 in the sun}

\author{ S.~Degl'Innocenti$^{1,2}$,
					 		G.~Fiorentini$^{1,3}$ and
          B.Ricci $^{1}$  }

\address{
$^{1}$Istituto Nazionale di Fisica Nucleare, Sezione di Ferrara,
      via Paradiso 12, I-44100 Ferrara, Italy\\
$^{2}$Dipartimento di Fisica dell'Universit\`a di Pisa,
       piazza Torricelli 1, I-56100 Pisa, Italy\\
$^{3}$Dipartimento di Fisica dell'Universit\`a di Ferrara,
      via Paradiso 12, I-44100 Ferrara, Italy
        }

\date{June 1997}
\maketitle                 

\begin{abstract}

By using a phenomenological field theory of nucleon-nucleon
interactions,  Oberhummer et al. found a cross section of
 p+p $\rightarrow$ d + e$^+$ + $\nu _e$
 about 2.9 times that given by the potential
approach and adopted in Standard Solar Model calculations.
We show that a solar model with  $S=2.9 S_{SSM}$ is inconsistent
with helioseismic data, the difference between model predictions
and helioseismic determinations being typically a factor ten larger
than estimated uncertainties. We also show that, according to
helioseismology, $\Sppzero$ cannot differ from $S_{SSM}$ by more than 15\%.

\end{abstract}

~\\
~\\

The rate of the initial reaction in the pp chain is too low to be
directly  measured in
the laboratory (even  in the solar  center this rate is extremely small,
of the order of 10$^{-10}$ yr$^{-1}$
consistently  with the  solar age) 
and it can be determined only by using the theory
of low energy weak interactions, together with the measured
properties of the deuteron and of the proton-proton scattering.
In terms of the astrophysical factor, $S$(E),
what really matters is its zero energy value, which for brevity
 will be indicated simply as $\Sppzero$. 
While we refer to \cite{BP92,KamB,Report} for  updated reviews, we remark that 
the calculated values 
 are all in the range  (3.89--4.21) 10$^{-25}$ MeV b,
i.e. they differ from their mean 
 by no more than 3\%.
In summary, as input of Standard Solar Model (SSM) calculations, one
takes \cite{KamB}:
\begin{equation}
\label{provo}
\Sppzero _ {SSM} =3.89 \cdot 10^{-25} (1 \pm 0.01) \, {\mbox{ MeV b}}  \quad .
\end{equation}

Although some warning is  
in order as to the meaning
of the quoted error, one may  conclude that well 
 known physics 
determines $\Sppzero$ to the level of few per cent or even better.

Recently, however, Oberhummer et al. \cite{Ober} presented 
a new evaluation of $\Sppzero$ in a relativistic field theory
framework, where strong interactions are phenomenologically
described by one nucleon loop diagrams. As well known since
\cite{Stein}, Adler-Bell-Jackiw anomalies are present in 
such models. The authors of Ref. \cite{Ober} 
claim that such anomalies provide 
the dominant contributions to the scattering amplitude, 
and that this effect has not be considered
in the potential approach, yielding to Eq. (\ref{provo}),
as being the result of a purely  field theory phenomenon without a 
classical analogue.
The estimated reaction rate is  a factor 2.9 times that of the
conventional approach.

Although it is not clear to us if the proposed field theory is suitable
for an accurate description of the low energy strong interaction
phenomenology (deuteron wave function, nucleon-nucleon scattering
amplitudes,...), it is of some interest to reconsider the effect
of varying $\Sppzero$ well beyond its estimated uncertainty, 
see Eq.(\ref{provo}).

The effect on the central solar temperature and on neutrino
fluxes has been discussed e.g.  in \cite{CDF93,BU89,BahUlm,Lebreton}.

As well known, and remarked in \cite{Ober}, a drastic
   increase of $S$ alleviates but does not solve the solar neutrino
   puzzle. The resulting low central temperature model implies a drastic
   reduction of $^7$Be neutrinos and an even stronger one for $^8$B neutrinos.
   As a consequence, the predicted Gallium signal is close
   to the observed values, but then Homestake, Kamiokande 
  and Superkamiokande
   are observing definitely too many ($^7$Be and/or  $^8$B)   neutrinos!

 In this
letter we shall concentrate on helioseismic implications of varying
$\Sppzero$.

Indeed, helioseismology allows us to look into the deep interior of the Sun, probably 
more efficiently than neutrinos (for reviews see 
\cite{Science,Oscill,ARAA,elios,c-d}). The highly  precise
 measurements of 
frequencies and  the tremendous number of measured lines enable us to extract 
the values of sound speed and density inside the Sun with accuracy better 
than $1\%$. 
Furthermore, from helioseismic data one derives accurate predictions on some
properties of the convective envelope: the transition of the 
temperature gradient between being subadiabatic and adiabatic at the 
base of the solar convective zone gives rise to a clear signature
in the sound speed \cite{elios}. Helioseismic measurements  therefore
determine the location $R_b$ and  the 
density $\rho_b$ of the base of the convective zone. 
In addition, the photospheric helium abundance $Y_{ph}$, which is of fundamental  importance
both to cosmology and to solar structure theory and which cannot be
determined by direct measurements, is constrained by helioseismology.
In Table \ref{tabh} we present the helioseismic determination of the 
above mentioned quantities $R_b, \, \rho_b$ and $Y_{ph}$, 
together with {\em conservative} estimates of the  uncertainties
due to both observational errors and inversion technique, see Ref. \cite{eliosnoi}.

Recent standard solar model calculations, including element
diffusion and using updated opacities and accurate equations of state, 
are well in agreement with helioseismic data, see Ref. \cite{eliosnoi}.
Let us compare with helioseismic data a solar model (hereafter MOD2.9)
 obtained from the 
FRANEC evolutionary code \cite{Ciacio} by taking $S=2.9 S _{SSM}$, all
other input parameters being kept at the SSM values.

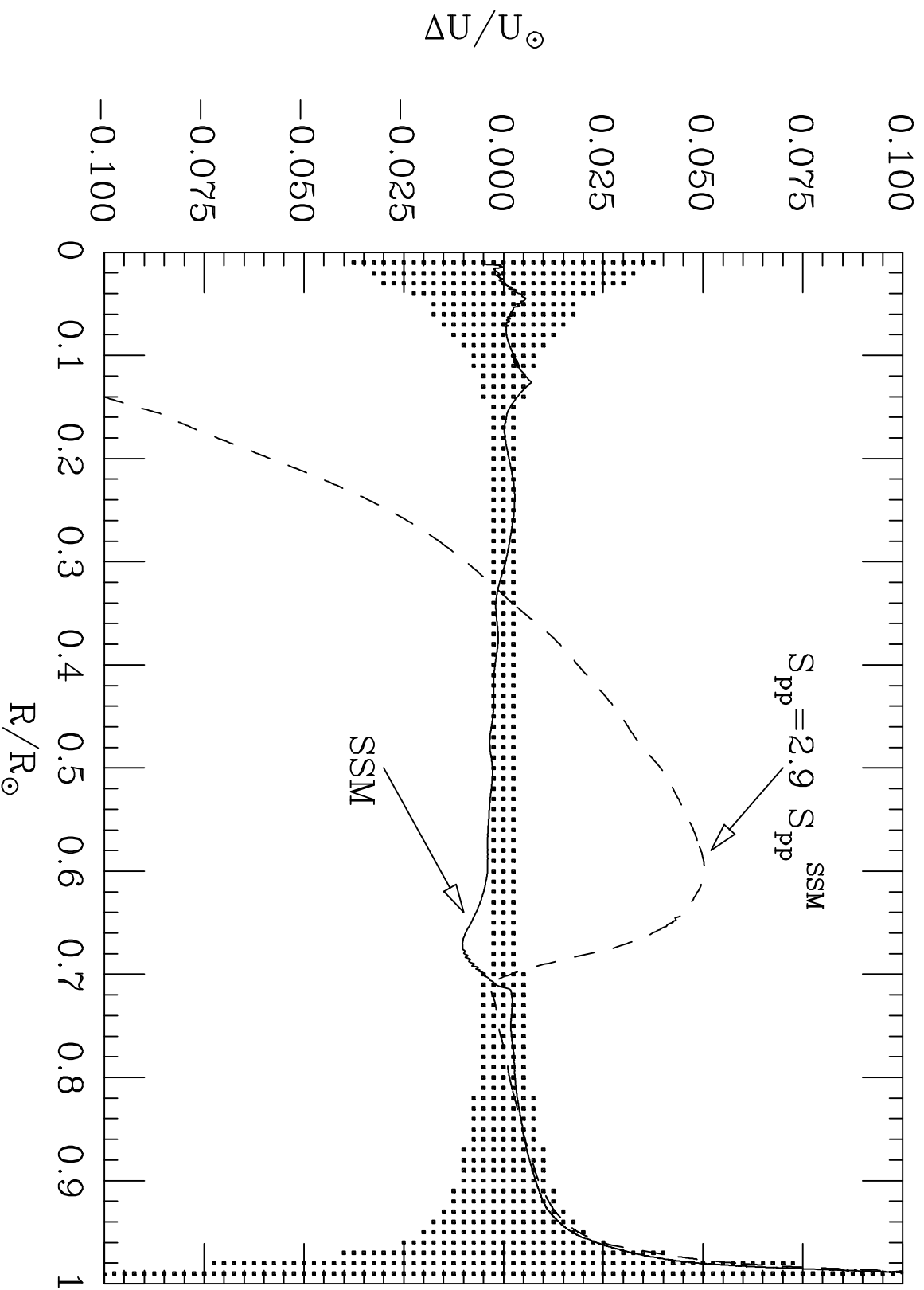
\begin{figure}
\caption[ff]{
The fractional difference $(U-U_\odot)/ U_\odot$ for the 
FRANEC-SSM (solid line) and for  the model with $S=2.9 S_{SSM}$
(dashed line). The dotted area corresponds to the uncertainty on $U_\odot$.
}
\label{figu}
\end{figure}

 Concerning the (isothermal) sound speed squared, $U=P/\rho$, 
the estimated accuracy of helioseismic determination is,
conservatively, of order 0.5\% for intermediate values
of the solar radial coordinate $R$. More precisely, 
as a function of $R/R_\odot$, the relative
accuracy of $U$ corresponds to the dotted area in Fig. \ref{figu}.
From the same figure one sees that the SSM satisfies
the helioseismic constraint almost everywhere, in that the error
band generally includes $(U_{SSM} - U_{\odot} )/ U_{\odot}$ 
where $U_{SSM}$ is the value predicted by the SSM and $U_{\odot}$
is the helioseismic determination.

On the other hand, for MOD2.9 the profile of 
 $(U_{2.9} - U_{\odot} )/ U_{\odot}$ 
looks clearly inconsistent with helioseismology. 
At $R\simeq 0.6 R_{\odot}$ the relative difference is of order
5\%, a factor ten beyond the estimated uncertainty of $U_\odot$
\footnote{We remark that sound speed profiles for $S\neq S_{SSM}$ have
 been discussed in Refs. \cite{TCL,Dzie}}.

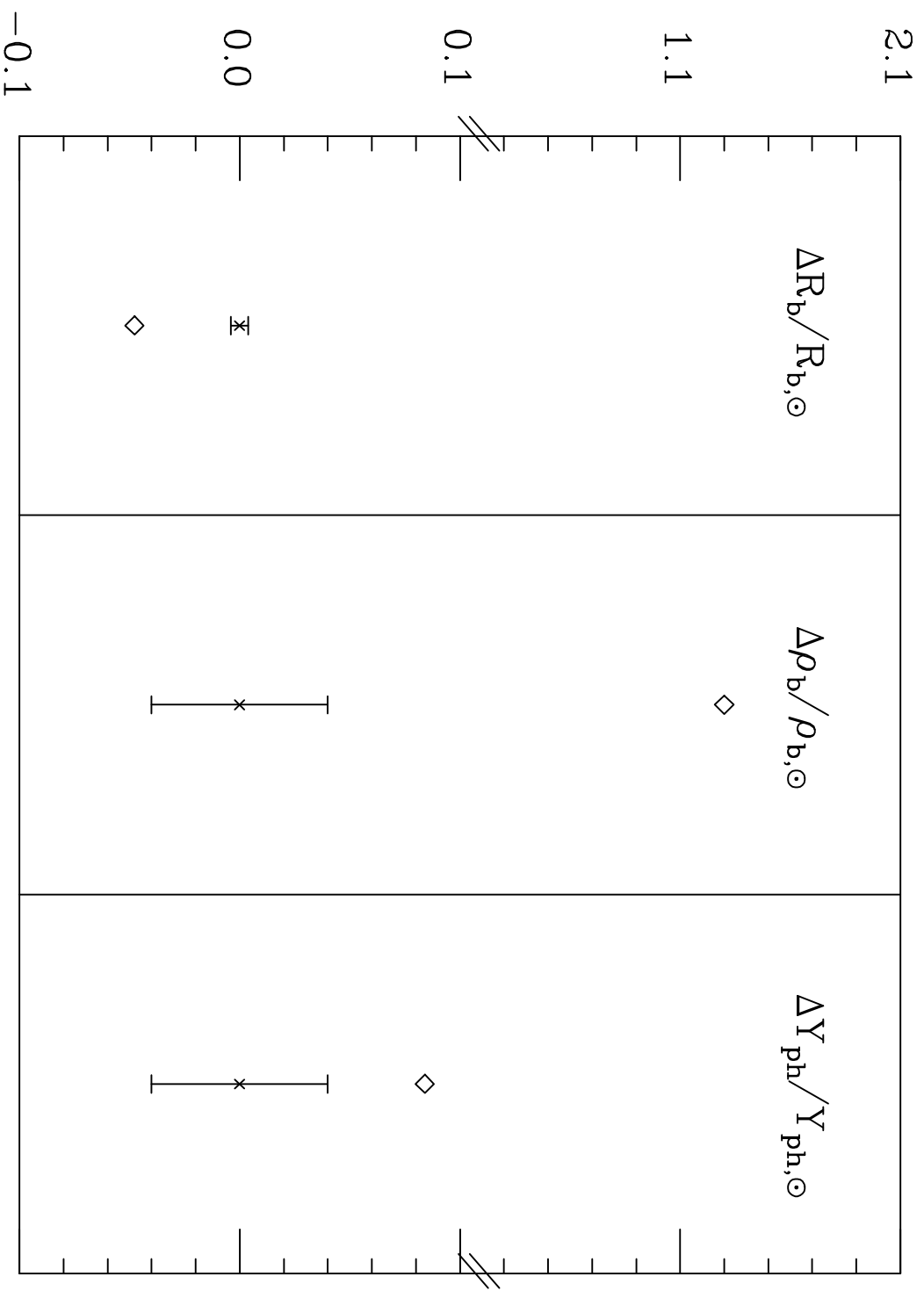
\begin{figure}
\caption[f]{
For the indicated quantities $Q$ we present the fractional difference
$(Q-Q_\odot)/ Q_\odot$ for  $S=2.9 S_{SSM}$ (diamonds) together
with the relative helioseismic uncertainties (bars).
}
\label{figb}
\end{figure}

 The comparison between the properties of the convective
envelope, see Table \ref{tabh} and Fig. \ref{figb}, also
shows the inadequacy of MOD2.9. For instance, the distance between the predicted
 and the true depth of the convective zone is ten times the
estimated error.

All this shows that $S=2.9 S_{SSM}$ is untenable.
On the other hand, we remind that only theoretical estimates
of $\Sppzero$ are available and observational
information would be welcome. In this respect, it is interesting to
determine the range of S-values which are acceptable in 
comparison with helioseismology.

We remind that there are two major uncertainties in building SSMs:
solar opacity $\kappa$  and heavy element abundance $\zeta =$Z/X
are only known with an accuracy of about $\pm$ 10\%.
By using $\kappa $ and $\zeta$ as free parameter within
their estimated uncertainties we can determine the acceptable range of
$\Sppzero$ as that such that $R_b,\, \rho_b$ and $Y_{ph}$ are
all predicted within the helioseismic range.

The dependence of these quantities on $\kappa$, $\zeta$ and $S$
has been determined numerically in Ref. \cite{eliostc}:
\begin{mathletters}
\begin{eqnarray}
\label{leggi2a}
R_b&=& R_{b, SSM} \, 
\left ( \frac{\kappa}{\kappa_{SSM}} \right )^{-0.0084} \,
\left ( \frac{\zeta}{\zeta_{SSM}} \right ) ^{-0.046} \,
\left ( \frac{S}{S_{SSM}} \right ) ^{-0.058} \\
\label{leggi2b}
\rho_b&=& \rho_{b,SSM} \, 
\left ( \frac{\kappa}{\kappa_{SSM}} \right )^{0.095} \,
\left ( \frac{\zeta}{\zeta_{SSM}} \right ) ^{0.47} \,
\left ( \frac{S}{S_{SSM}} \right ) ^{0.86} \\
\label{leggi2c}
Y_{ph}&=& Y_{ph, SSM} \, 
\left ( \frac{\kappa}{\kappa_{SSM}} \right )^{0.61} \,
\left ( \frac{\zeta}{\zeta_{SSM}} \right ) ^{0.31} \,
\left ( \frac{S}{S_{SSM}} \right ) ^{0.14} 
\end{eqnarray}
\end{mathletters}
Most of the information on $S$ arises from data on $\rho_b$
as this observable depends strongly on $S$ whereas it is
weakly affected by the others parameters.
One can understand the dependence on $S$, at least qualitatively.
A value of $S$ larger than $S_{SSM}$ implies smaller temperature in the
solar interior, which thus becomes more opaque
 (in other words, the region of partial ionization is deeper).
 Radiative transport therefore
is less efficient and convection starts deeper
in the Sun ($R_b < R_{b,SSM}$) where density is higher ($\rho_b >\rho_{b,SSM}$).

By using Eqs. (2) and the allowed ranges  reported in Table \ref{tabh},
also taking into account the predictions of different SSMs, we find:
\begin{equation}
0.94 \leq  S/S_{SSM} \leq 1.18
\end{equation}

In conclusion, we remark that helioseismology provides the only
observational constraint, although indirect, on the
 p+p $\rightarrow$ d + e$^+$ + $\nu _e$ reaction.

\acknowledgments
We are extremely grateful to  V. Castellani,  W.A. Dziembowski 
and M. Moretti
  for fruitful comments and suggestions.

\begin{table}
\caption[abc]{
For the depth of the convective zone $R_b$,
the density at the bottom of the convective zone $\rho_b$ 
and the photospheric helium abundance $Y_{ph}$ we
present the helioseismic determination, from Ref. \cite{eliosnoi},
and the predictions corresponding to $S=2.9 S_{SSM}$.
}
\begin{tabular}{c c c c c }
&Q & Helioseismology &  MOD2.9 \\
\hline
&$R_b/R_\odot$  & 0.711 $\pm 0.003$ & 0.677 &\\
&$\rho_b$ [g/cm$^3$] &  0.192  $\pm 0.007 $ & 0.409 &\\
&$Y_{ph}$ &   0.249 $\pm 0.011$ & 0.270  & \\
\end{tabular}
\label{tabh}
\end{table}

\end{document}